\documentclass[aps,pra,twocolumn,showpacs,floatfix,superscriptaddress]{revtex4}

\usepackage{dcolumn}                 
\usepackage{times}
\usepackage{nicefrac}
\usepackage{amsmath}
\usepackage{amsfonts}
\usepackage{amssymb}
\usepackage{amsthm}
\usepackage{bm}

\usepackage{color}
\usepackage{graphics}
\usepackage{graphicx}
\usepackage{epsfig}

\usepackage{dcolumn}                 
\newcolumntype{w}[1]{D{.}{.}{#1}}

\newcommand*{\cent}[1]{\multicolumn{1}{c}{$#1$}}

\newcommand{\dd}{\mathrm{d}}
\newcommand{\ee}{\mathrm{e}}
\newcommand{\kHz}{\mathrm{kHz}}
\newcommand{\Hz}{\mathrm{Hz}}
\newcommand{\KK}{\mathrm{K}}

\newcommand{\addrMissouri}{Missouri University of Science and
Technology, Rolla, Missouri 65409-0640, USA}

\newcommand{\addrGaithersburg}{National Institute of Standards and Technology,
Gaithersburg, Maryland 20899-8420, USA}

\newcommand{\addrPoznan}{Faculty of Chemistry,
Adam Mickiewicz University, Grunwaldzka 6, 60-780 Pozna\'n, Poland}

\begin{document}

\title{Black--Body Radiation Correction to the Polarizability of Helium}

\author{M. Puchalski}
\affiliation{\addrMissouri}
\affiliation{\addrPoznan}

\author{U. D. Jentschura}
\affiliation{\addrMissouri}

\author{P. J. Mohr}
\affiliation{\addrGaithersburg}

\date{\today}

\begin{abstract}
The correction to the polarizability of helium due to black-body radiation is
calculated near room temperature.  A precise theoretical determination of the
black-body radiation correction to the polarizability of helium is essential
for dielectric gas thermometry and for the determination of the Boltzmann
constant. We find that the correction, for not too high temperature, is roughly
proportional to a modified hyperpolarizability (two-color hyperpolarizability),
which is different from the ordinary hyperpolarizability of helium.  Our
explicit calculations provide a definite numerical result for the effect and
indicate that the effect of black-body radiation can be excluded as a limiting
factor for dielectric gas thermometry using helium or argon.
\end{abstract}

\maketitle

%
%
\section{Introduction}

The recent measurement~\cite{ScGaMaMo2007} of the refractive
index of helium in a microwave cavity resonator has yielded
the hitherto most precise experimental value of the 
molar polarizability of the helium atom,
\begin{equation}
A_\epsilon = \frac{ \alpha_d \, N_A }{3 \, \epsilon_0}
           = 0.5172535(47) \, \frac{{\rm cm}^3}{{\rm mol}} \,.
\end{equation}
Here, $N_A$ is Avogadro's number, $\epsilon_0$ is the 
vacuum permittivity (also called the ``electric constant''), and
$\alpha_d$ is the static electric dipole polarizability of helium.  We also
recall that $A_\epsilon$ is defined in the limit of zero density.  The
techniques in this experiment could lead to measurements of thermodynamic
temperature or to a determination of the value of the Boltzmann constant. Each
of these applications would take advantage of the fact that $\alpha_d$ has been
accurately determined theoretically in a series of calculations including
complete leading relativistic $\alpha^2$ and
quantum electrodynamics (QED) $\alpha^3$ corrections in the
fine structure constant expansion \cite{LuGrFe1996,PaSa2000,CeSzJe2001,LaJeSz2004} with an
uncertainty of 0.2~ppm from the estimate of the $\alpha^4 $ term.  Here, the
magnitude of the corrections is given in atomic units, i.e., relative to the
Hartree energy scale.
We note that for excitation by low-energy radiation,
as is relevant for the experiment~\cite{ScGaMaMo2007},
the relativistic and radiative
corrections to the polarizability are unambiguously defined.
However, for higher frequencies, there may be additional
field-configuration dependent corrections (see Appendix E of Ref.~\cite{HaEtAl2006}).

Black-body radiation present in the cavity resonator will lead to a
temperature-dependent correction to the measured value of the helium molar
polarizability, which could affect the interpretation of such measurements.
However, the correction due to black-body radiation was assumed to be
negligible compared to the uncertainty of the measurement in Ref.
\cite{ScGaMaMo2007}.
(Heuristic arguments that support this assumption and
indicate that it is also valid for argon are presented in
Sec.~\ref{conclu} below.)
The experiment~\cite{ScGaMaMo2007} determined the
polarizability of helium atoms through the interaction of microwaves with the
atoms in a cavity resonator. At the same time, black-body radiation is present
in the resonator, and it also interacts with the atoms. As a first
approximation, the interaction of the cavity microwaves with the atoms and the
interaction of the black-body radiation with the atoms are independent
processes and do not affect each other.  The dominant energy shifts due to
microwave radiation on the one hand, and due to black-body radiation on the other
hand, are described by the corresponding second-order AC Stark
shifts~\cite{Sa1994Mod,JeHa2008,PoDe2006}
and are proportional, to very good approximation, to
the dynamic second-order dipole polarizability at the microwave and black-body
frequencies.  The derivation of the theoretical second-order expressions in
both classical, time-ordered perturbation theory and in the field-quantized
framework are contrasted against each other in Ref.~\cite{HaJeKe2006}.  From a
quantum electrodynamic (QED) point of view, both photon annihilation as well as
photon creation processes contribute to the dynamic polarizability.

Besides the second-order polarizability, there is also a
fourth-order effect which is due to the exchange of four instead of two
photons with the radiation field(s). When the
radiation is monochromatic, the total fourth-order
energy shift is proportional to the so-called hyperpolarizability
of the atom~\cite{GrChHu1968}.
However, when the atom is simultaneously interacting
with both microwave as well as black-body radiation, the
treatment has to be modified because photon creation and annihilation
processes of one and the same field have to be ``matched,''  and
this excludes some intermediate, virtual states of the
atom$+$radiation field from the fourth-order expressions.
Indeed, in generalizing the fully quantized formalism to fourth order,
we find convenient expressions which describe the two-color
hyperpolarizability.
The resulting fourth-order energy shift finds a natural interpretation
as a perturbation of the second-order dynamic energy shift due
to the microwave photons. The latter is proportional to the
dynamic polarizability. Therefore, the fourth-order effect
constitutes a correction to the dynamic polarizability of the atom.

Note that some of the QED
corrections to the polarizability of helium also
involve fourth-order perturbation theory~\cite{PaSa2000,LaJeSz2004},
with black-body photon interactions being replaced by the interactions
with the radiative photons. However, there is an important
difference. E.g., for the QED corrections to the Bethe logarithm,
the atom only emits then absorbs virtual radiative photons,
while it emits then absorbs, and absorbs then emits photons with the
probing electromagnetic waves. In the evaluation of the
black-body radiation correction to the dynamic polarizability,
we have to take into account processes where the atom
both emits then absorbs and absorbs then emits black-body {\em and}
probing microwave photons.

Our paper is organized as follows.
The theoretical foundations are recalled and derived in Sec.~\ref{theory}.
Numerical investigations are described in Sec.~\ref{numerical}.
Conclusions are reserved for Sec.~\ref{conclu}.

%
%
\section{Theory}
\label{theory}

In order to formulate the problem, we need to take into account the interaction
of the helium atom with two electromagnetic fields: (i) the microwave
field used to probe the electric dipole polarizability, and (ii) the
black-body radiation field.
We work with a second quantized radiation field and
with a first quantized theory for the atomic electrons.
Furthermore, we work in the Schr\"odinger picture.
In SI units, the Hamiltonian
for the helium atom coupled to an
external source of microwaves and affected by black-body radiation
is given as
\begin{equation}
\label{H}
H = H_0 + \hbar \omega_M a_M^\dagger \, a_M
+ \sum_B \hbar \omega_B a_B^\dagger a_B
+ {\cal U}_M + {\cal U_B}.
\end{equation}
Here, $M$ and $B$ are multi-indices defined as
\begin{equation}
\label{multi}
M \equiv \vec{k}_M \lambda_M, \qquad
B \equiv \vec{k}_B \lambda_B,
\end{equation}
where $\vec{k}_M$ and $\vec{k}_B$ are the wave vectors of the
probing microwave field and of the black-body field,
and $\lambda_{M,B}$ denote their polarizations.
We sum over the modes of the black-body field
and assume that the microwave radiation can be described by a
single mode. The other symbols used in Eq.~\eqref{H} are as follows.
In the nonrecoil approximation,
the helium atom is described by the nonrelativistic Hamiltonian
\begin{equation}
H_0 = \sum_{a=1,2} \left( \frac{\vec p_a^{\,2}}{2 m}
- \frac{Z e^2}{4 \pi \epsilon_0 r_a}\right)
+ \frac{e^2}{4 \pi \epsilon_0 r_{12}} \,,
\end{equation}
where $r_{12}$ is the distance between the electrons, $e$ is the electron
charge, $Z = 2$ the nuclear charge number, $\epsilon_0$ is the vacuum
permittivity, and $r_a$ ($a = 1,2$) is the electron-nucleus distance. We
describe the interaction of the atom with the quantized electromagnetic fields
in the length gauge.  The electric dipole interactions of the atom with the
microwave and black-body fields are as follows,
\begin{eqnarray}
{\cal U}_M &=&
- e \sqrt{ \frac{\hbar \omega_M}{2 \epsilon_0 {\cal V}_M}} \;
\vec \epsilon_M \cdot \vec{r} \;
\left( a_M^{\dagger} + a_M \right)\,, \\
{\cal U}_B &=& -e \, \sum_B
\sqrt{\frac{\hbar \omega_B}{2 \epsilon_0 {\cal V}_B}}  \;
\vec \epsilon_B \cdot \vec{r} \;
\left( a_B^{\dagger} + a_B \right) \,,
\label{df_VLVB}
\end{eqnarray}
where the photon creation and annihilation operators
are $a^\dagger$ and $a$, respectively.
We normalize the electric field operators
(see Ref.~\cite{JeKe2004aop}) so that the energy density of the
microwave photon integrated over the volume ${\cal V}_M$ is equal to
$\hbar \omega_M$, and analogously for the black-body photon. The effect of
the electromagnetic fields is assumed to be a small perturbation
of $H_0$, so that a perturbative treatment of the
dipole interaction of the atom with the electromagnetic field
becomes possible.

We first consider the second-order effect which gives the
main energy perturbation of the helium atom due to the
AC Stark effect. A single-mode microwave field probes the helium
atom in the ground state.
The ground state energy is denoted as $E_0$ and and its
Schr\"{o}dinger wave function is denoted as $\psi_0$. First-order perturbation
theory in ${\cal U}_M$ gives a vanishing effect, and the
leading correction to the energy $E_0$ is of second order.
We then average over the polarizations and propagation
directions of the microwave mode. We are interested in the
classical limit where the number of microwave photons is $n_M \gg 1$,
the normalization volume is large
(${\cal V}_M \gg 1$), but the ratio $n_M /{\cal V}_M$
remains finite,
and proportional to the intensity of the microwave field.
Using this formalism,
one may easily rederive~\cite{HaJeKe2006}
the dynamic AC Stark energy shift due to microwave photons,
\begin{eqnarray}
\label{pol}
\Delta E &=& \  \frac{e^2}{3} \left( \frac{n_M \hbar \omega_M}{2
\epsilon_0 {\cal V}} \right) \sum_{\pm} \langle \psi_0 |  r^i {\cal
R}(\pm \omega)  r^i  | \psi_0 \rangle \,,
\end{eqnarray}
where
\begin{equation}
{\cal R}(\pm \omega) = \frac{1}{E_0 - H_0 \mp \hbar \omega} =
\sum_{m} \frac{|\phi_m \rangle \langle \phi_m|}{E_0 - E_m \mp \hbar \omega}
\end{equation}
is the resolvent operator for the unperturbed helium atom, and $r^i
= r_1^i + r_2^i$; that is, we are denoting the $i$th
Cartesian component of the sum $\vec r_1 + \vec r_2$ of the
positions $\vec r_1$ and $\vec r_2$ of both electrons simply by
$r^i$. For repeated superscripts and subscripts, 
we use the summation convention
[an example is given by 
the Cartesian superscripts $i$ in Eq.(\ref{pol})]. The sum over
$\pm$ is necessary because we treat both photon absorption followed
by emission as well as photon emission followed by absorption. The
intensity $I_M$ of the microwave field in our normalization of the
field operator is given as
\begin{equation}
I_M = \frac{n_M \hbar \omega_M c}{{\cal V}_M} \,.
\end{equation}
For our purposes (ground state of helium perturbed
by a microwave field), we may approximate to good accuracy
the microwave frequency by the static limit
($\omega = \omega_M \to 0$) in Eq.~\eqref{pol}.
Then, the well-known final result in second order is
rederived,
\begin{equation}
\label{df_EL}
\Delta E = -\frac{e^2 I_M}{2 \epsilon_0 \, c} \alpha_d \,,
\end{equation}
where the static dipole polarizability 
(divided by the square of the
elementary charge $e^2$)
in nonrelativistic limit for the ground state
is defined by
\begin{equation}
\label{df_alphad}
\alpha_d =
-\frac{2}{3} \langle \psi_0 |  r^i  {\cal R}'(0) r^i | \psi_0 \rangle\,.
\end{equation}
Here, ${\cal R}'(0) = 1/(H_0 - E_0)'$ is the reduced Green function,
where the reference state $| \psi_0 \rangle$ is
excluded from the sum over intermediate (virtual) states.

We now investigate the perturbation of the dipole polarizability
$\alpha_d$ due to the black-body radiation. This is a fourth-order
process in the radiation field. The atom may emit then absorb photons
from the microwave field and also emit then absorb photons from the
black-body field. Fourth-order perturbation theory, with
time-independent field operators in the Schr\"{o}dinger picture, can
then be used in order to infer the energy shift. The result reads,
when taking all combinations of emission and absorption into
account, 
%
\begin{widetext}
\begin{align}
\label{df_gomega} & \Delta E_{B} = e^4 \left( \frac{\hbar
\omega_M}{2 \epsilon_0 {\cal V}_M}\right) \sum_B \left( \frac{\hbar
\omega_B}{2 \epsilon_0 {\cal V}_B}\right) \; \biggl\{ 2  \, \langle
(\Xi_B\, a^{\dagger}_B \; {\cal R}(\omega_M) \, \Xi_M \,
a_M^{\dagger} {\cal R}(\omega_B + \omega_M) \, \Xi_B \, a_B {\cal
R}(\omega_M)
\Xi_M a_M \rangle \nonumber \\
&+ \langle
\Xi_B a^{\dagger}_B  {\cal R}(\omega_B) \Xi_M a_M^{\dagger}
{\cal R}(\omega_B + \omega_M) \Xi_M a_M
{\cal R}(\omega_B) \Xi_B a_B \rangle
+ \langle \Xi_M a_M^{\dagger} {\cal R}(\omega_M) \Xi_B a^{\dagger}_B
{\cal R}(\omega_B + \omega_M) \Xi_B a_B
{\cal R}(\omega_M) \Xi_M a_M \rangle \nonumber \\[2ex]
&+ 2 \, \langle \Xi_B a^{\dagger}_B
{\cal R}(\omega_B) \; \Xi_B a_B  \;
{\cal R}'(0) \; \Xi_M a_M^{\dagger}  \;
{\cal R}(\omega_M) \; \Xi_M a_M  \rangle
- \langle \Xi_B a^{\dagger}_B \;
{\cal R}(\omega_B) \; \Xi_B a_B \rangle \;
\langle \Xi_M a_M^{\dagger} \;
{\cal R}^2(\omega) \Xi_M a_M \rangle
\nonumber \\
& - \langle \Xi_B a^{\dagger}_B \;
{\cal R}^2(\omega_B) \; \Xi_B a_B \rangle
\langle ( \Xi_M a_M^{\dagger} \;
{\cal R}(\omega_M) \; \Xi_M  a_M \rangle \biggr\}
+ \left< (a^{\dagger}_B, -\omega_B) \leftrightarrow (a_B, \omega_B) \right>
+ \left< (a^{\dagger}_M, -\omega_M) \leftrightarrow (a_M, \omega_M) \right>
\nonumber \\
& + \left< (a^{\dagger}_B, -\omega_B) \leftrightarrow (a_B, \omega_B);\;\;
(a^{\dagger}_M, -\omega_M) \leftrightarrow (a_M, \omega_M) \right> \,,
\end{align}
\end{widetext}
%
where we use the shorthand notation
\begin{equation}
\Xi_B \equiv \vec\epsilon_B \cdot \vec r \,, \qquad
\Xi_M \equiv \vec\epsilon_M \cdot \vec r \,.
\end{equation}
Furthermore, we again use multi-indices $M$ and $B$
as defined in Eq.~\eqref{multi}.
The replacement terms in Eq.~\eqref{df_gomega}
correspond to the exchange of black-body photon
emission versus absorption, of
microwave photon emission versus absorption,
and simultaneous exchange of both processes.
Next, we consider the ``classical limit''
of a high occupation number for both fields,
and the low frequency limit for the microwave field,
$\omega_M \to 0$ (this approximation is always valid for microwave photons
whose energy is low compared to the first available
atomic transition). Furthermore, we match the summation over the
black-body modes $B$ with an integration over the frequency-dependent
intensity of the black-body radiation in Eq.~\eqref{df_gomega},
\begin{equation}
\sum_B \frac{n_B \hbar \omega_B c}{{\cal V}_B} \rightarrow
\int_0^{\infty} \dd \omega_B \, u(\omega_B, T) \,,
\end{equation}
where Planck's  law gives
\begin{equation}
u(\omega_B ,T) \, \dd \omega =
\frac{\hbar }{4 \pi^3 c^2} \;
\frac{\omega_B^3 \; \dd \omega_B}{\exp (\hbar \omega_B/ k_B T) - 1}.
\end{equation}
For room temperature, the black-body spectrum has its maximum at
frequencies much below typical atomic transition frequencies.
Therefore, in addition to approximating the microwave frequency in
Eq.~\eqref{df_gomega} by zero, we may also approximate the
black-body frequency by zero in the fourth-order polarizability
defined in Eq.~\eqref{df_gomega}.
Indeed, if we employ the approximation
$\omega_B \to 0$ in the matrix element defined in
Eq.~\eqref{df_gomega} (not in the prefactor which
is proportional to $\omega_B$), then we may even integrate over the
black-body photon frequency analytically.
This is explored
in the following. For now, we approach the problem
by numerically integrating over the
black-body photon frequency.  The result can be written as
\begin{equation}
\label{defchi}
\Delta E_{B} =  \frac{e^2 I_M}{2 \epsilon_0 c} \alpha_d \; \chi =
\Delta E \; \chi\,,
\end{equation}
with the dimensionless factor
\begin{equation}
\chi = \int_0^{\infty} \dd\omega_B \; \chi(\omega_B,T) \,.
\end{equation}
Thus, the product $\alpha_d \, \chi$ can be viewed as an
effective static dipole polarizability of helium in the presence of
black-body radiation where $\alpha_d$ is the 
(dipole) polarizability in the
absence of the black-body radiation, and $\chi$ is a multiplicative
factor that gives the change in the measured value due to the
radiation.
The integrand $\chi(\omega_B,T)$ involves the
thermal distribution of photons,
\begin{equation}
\label{df_chioB}
\chi(\omega_B,T) = \frac{4}{9}
\frac{e^2}{\alpha_d\epsilon_0 c}\sum_{\pm} g(\pm \omega_B) \;
u(\omega_B, T) \,.
\end{equation}
The $g$ function is obtained after summing over the photon
modes in Eq.~\eqref{df_gomega} and after taking into
account all photon creation and annihilation processes, and reads
\begin{widetext}
\begin{align}
g(\omega_B)  \equiv & \;
- \langle \psi_0 | r^j {\cal R}(\omega_B) r^k {\cal R}(\omega_B) r^j {\cal R}'(0) r^k
| \psi_0 \rangle
- \langle \psi_0 | r^j {\cal R}(\omega_B) r^j {\cal R}'(0) r^k {\cal R}'(0) r^k
| \psi_0 \rangle
\nonumber \\
& \; - \frac{1}{2}\ \biggl( \langle \psi_0 | r^j {\cal R}(\omega_B)
r^k {\cal R}(\omega_B) r^k {\cal R}(\omega_B) r^j | \psi_0 \rangle
+ \langle \psi_0 | r^j {\cal R}'(0) r^k {\cal R}(\omega_B) r^k {\cal R}'(0)
r^j  | \psi_0 \rangle \nonumber \\
& \; - \langle \psi_0 | r^j {\cal R}(\omega_B) r^j | \psi_0 \rangle \;
\langle \psi_0 | r^k {\cal R}'^{2}(0) r^k  | \psi_0 \rangle
- \langle \psi_0 | r^j {\cal R}^2(\omega_B) r^j | \psi_0 \rangle \;
\langle \psi_0 |  r^k {\cal R}'(0)  r^k  | \psi_0 \rangle \biggr) \,.
\label{df_g}
\end{align}
\end{widetext}
The dimensionless
factor $\chi$ is the quantity we are looking for as it gives the relative
perturbation $\chi$ to the polarizability
due to black-body radiation on $\alpha_d$ according to Eq.~\eqref{defchi}.

To this point, we have kept SI MKSA units in all formulas,
as practiced by the Committee on Data for Science and Technology (CODATA).
In calculations of
atomic properties, it is usually more convenient to use formulas in atomic
units. The SI MKSA static polarizability $\alpha_d$, the
angular frequency $\omega_B$ and $g(\omega_B)$ from Eq.~(\ref{df_g}) are
related to their atomic unit counterparts
$\overline\alpha_d$, $\overline\omega_B$, and $ \overline g(\overline
\omega_B)$ by
\begin{subequations}
\label{atu}
\begin{eqnarray}
\alpha_d &=& \overline \alpha_d \frac{a_0^2}{E_h} \,, \\
\omega_B &=& \frac{E_h}{\hbar} \, \overline \omega_B \,, \\
g(\omega_B) &=& \overline g(\overline \omega_B) \,
\frac{a_0^4}{E_h^3}\,,
\label{df_atunits}
\end{eqnarray}
\end{subequations}
where $a_0$ is the Bohr radius and $E_h$ is the
Hartree energy.  Then, $\chi$ as defined in Eq.~\eqref{defchi}
can be obtained as
\begin{align}
\label{df_chia}
& \chi = \int_0^\infty \dd \overline \omega_B \;
\overline\chi(\overline\omega_B, T) = \frac{4}{9}
\frac{\alpha^3}{\pi^2}
\nonumber\\[2ex]
& \; \times \sum_{\pm}
\int_0^\infty
\frac{\overline g(\pm \overline\omega_B)}{\overline \alpha_d}\,
\left[ \exp\left( \dfrac{E_h}{ k_B \, T} \, \overline\omega_B\right) - 1
\right]^{-1} \,
\overline\omega_B^3 \, \dd\overline\omega_B \,.
\end{align}
The atomic unit system is defined so that physical quantities pertaining to
atoms are of order one.  We can thus conclude that $\chi$ is an effect of order
$\alpha^3$ which is additionally suppressed by the Boltzmann factor.  In the
following, we use atomic units (i.e., units with $e = \hbar = m_e = 1$, and
where the length is measured in Bohr radii).

%
%

\section{Numerical calculation}
\label{numerical}

The crucial step in the evaluation is the calculation of the $g$ function
defined in Eq.~\eqref{df_g}.  According to Eq.~\eqref{df_g}, it contains both
manifestly fourth-order but also products of second-order terms. Next, we
integrate $g$ over $\omega_B$ in the interval $(0,\infty)$ with a weight given
by the Boltzmann distribution of the black-body radiation. The numerical
integration is not completely trivial, because we have to omit poles due to
resonances given by the virtual state in the denominators of the resolvent
${\cal R}( \omega_B)$.
By contrast, ${\cal R}(-\omega_B)$ has no poles.
Bending the integration contour into the complex plane
around the resonances solves the problem.  Moreover, all intermediate discrete
states with a positive factor $1/( E_0 -  E_m \pm  \omega_B) > 0$ must be
represented very accurately as they define the position of the resonances.

In practice, we are interested in temperatures that do not exceed the room
temperature substantially ($< 400\,{\rm K}$).  This simplifies the problem,
because the weight given by the Boltzmann factor is
exponentially suppressed on the scale of atomic transition frequencies. For $T
= 273\,\KK$, the maximum of the black-body radiation distribution lies at
$\omega_{B\mathrm{max}} \approx 0.00244 \, \rm{a.u.}$ (atomic units).
If we are interested in
evaluating the total effect to a relative accuracy of 1$\%$, we may
cut off the integration interval at $\omega_{B\mathrm{cut}} = 0.05$ a.u.,
where $u(\omega_{B\mathrm{cut}}, 273\,\KK)  /
u(\omega_{B\mathrm{max}},273\,\KK) \sim 10^{-20}$, and the ratio is even
smaller for lower temperatures. The function $g$ varies only marginally in the
frequency range relevant to the black-body radiation at 273K, and
$\omega_{B\mathrm{cut}}$ is still an order of magnitude less than the
energy (frequency) difference of the $1S$ ground state of helium
and the lowest excited states.  On the integration path from zero to
$\omega_{B\mathrm{cut}}$, we never approach the singular points in the
denominators of the resolvents in Eq.~\eqref{df_g}, and thus
only real values of $\omega_B$ need to be considered.

The nonrelativistic wave function of the ground state  $\psi_0$ and
its energy $ E_0$ in atomic units are determined based on the Rayleigh-Ritz
variational principle. We use a basis set of explicitly
exponentially correlated functions (see Refs.~\cite{Ko2000,Ko2002}
and also Appendix~\ref{appa})
\begin{equation}
\psi(r_1,r_2,r_{12}) =
\sum_{m=1}^{N_S} v_m \bigr[ \ee^{-a_i r_1 - b_i r_2 - c_i r_{12} } -
(r_1 \leftrightarrow r_2) \bigl] \,,
\label{df_psi0}
\end{equation}
where the parameters $(a,b,c)$ for the $i$th function are randomly generated
from an optimized box $(A_1,A_2) \times (B_1,B_2)\times (C_1,C_2)$ under
additional constraints $a_i + b_i > \varepsilon$ as well as $b_i + c_i >
\varepsilon$ and $c_i + a_i > \varepsilon$, where $\varepsilon = \sqrt{2 \,
\left( E_0^{+} -  E_0 \right)}$ and $E_0^{+}$ is
the ground state energy for He$^{+}$.  
In atomic units, $\varepsilon$ can be interpreted as an
approximate radial momentum of the two-electron system that characterizes
the radial exponential fall-off of the wave function. The
minimal momentum $\varepsilon$ must be chosen
to be large enough to be consistent with the binding of the electrons to the
helium nucleus. By requiring that all combinations $a_i + b_i$, $b_i + c_i$,
and $a_i + c_i$ fulfill this criterion,
we ensure that the wave function 
falls off sufficiently rapidly at large $r_1$, $r_2$, and $r_{12}$.
If a randomly generated orbital fails
to fulfill the requirements, we generate another one until conditions are met.
This method follows ideas outlined in Refs.~\cite{Ko2000,Ko2002}.  In order to
fix ideas, we should reemphasize that the six boundary parameters
characterizing the box, that is, $A_1$, $A_2$, $B_1$, $B_2$ $C_1$ and $C_2$ are
subject to variational optimization, not the random parameters $a_i$, $b_i$ ad
$c_i$.

In order to obtain a more accurate representation of the wave
function, we use two boxes that model the short-range and
medium-range asymptotics of the helium wave functions. For the
calculation of the fourth-order effect which is the subject of this
paper, matrices with a moderate number of $2 N_S=100$, $200$, $300$,
$400$ and $600$ basis functions are fully sufficient (we use a
prefactor $2$ in order to clarify the---equal---distribution of the
basis functions onto the two boxes).

In this basis, all needed matrix
elements can be represented as linear combinations of the integrals
(see Appendix~\ref{appa})
\begin{align}
& \Gamma(a,b,c,n_1,n_2,n_{12}) \nonumber\\[2ex]
& = \int \dd^3 r_1 \; \dd^3 r_2 \;
r_1^{n_1 - 1} r_2^{n_2 - 1} r_{12}^{n_{12}- 1} \,
\ee^{-a r_1 - b r_2- cr_{12}}
\end{align}
with nonnegative integers $n_1$, $n_2$, and $n_{12}$.  Recurrence relations
for their computation are well known \cite{SaRoKo1967,Ko2002}.
The result for the ground-state energy
extrapolated from 600 functions is $ E_0 = -2.903\,724\,377\,034\,118(3)$.
The linear coefficients $v_i$ in Eq. (\ref{df_psi0}) are obtained from
a solution of the generalized eigenvalue problem. The
numerical accuracy of the results is estimated
from the apparent numerical convergence of the matrix elements
as the size $N_S$ of the $S$ state basis is increased.

\begin{table*}[htdp]
\caption{\label{table1} Various quantities of
interest for the calculation of the
two-color hyperpolarizability, expressed in atomic units.
The definition of $\alpha_P$, $\alpha_{PP}$, $\alpha_D$
and $\alpha_S$ is given in Eqs.~\eqref{alphaP},
\eqref{alphaPP}, and \eqref{alphaDS}.
Numerical values are also indicated for the two-color
hyperpolarizability $\gamma_2 \equiv g(\omega = 0)$
defined in Eq.~\eqref{df_g0}, and for the
hyperpolarizability $\gamma$ defined in Eq.~\eqref{defgamma}.}
\begin{center}
\begin{tabular}{lllllll}
\hline
\hline
$2 N_S$ &
\cent{ \alpha_P} &
\cent{ \alpha_{PP}} &
\cent{ \alpha_S} &
\cent{ \alpha_D} &
\cent{ \gamma} &
\cent{ \gamma_2 \equiv g(0)}\\
\hline \\
 100 & -2.074\,787\,392\,227\,2 & 2.122\,527\,900\,595 & -11.261\,152\,615
     & -7.755\,245\,582  & 43.103\,075\,16 & 59.750\,569\,45 \\
 200 & -2.074\,788\,259\,836\,5 & 2.122\,530\,424\,544 & -11.261\,406\,459
     & -7.755\,398\,231  & 43.104\,221\,68 & 59.752\,012\,04 \\
 300 & -2.074\,788\,260\,731\,2 & 2.122\,530\,428\,743 & -11.261\,407\,099
     & -7.755\,398\,608  & 43.104\,224\,56 & 59.752\,015\,66 \\
 400 & -2.074\,788\,261\,670\,8 & 2.122\,530\,432\,055 & -11.261\,407\,836
     & -7.755\,399\,056  & 43.104\,227\,94 & 59.752\,019\,89 \\
 600 & -2.074\,788\,261\,679\,1 & 2.122\,530\,432\,021 & -11.261\,407\,802
     & -7.755\,399\,033  & 43.104\,227\,78 & 59.752\,019\,68 \\
\hline
\rule[-3mm]{0mm}{8mm}
$\infty$
     & -2.074\,788\,261\,682(3) & 2.122\,530\,432\,01(2) & -11.261\,407\,80(2)
     & -7.755\,399\,03(2) & 43.104\,227\,7(1) & 59.752\,019\,7(1) \\
\hline
Literature
     & -2.074\,788\,261\,682\,(3)$^a$ &    &  &  &  43.104\,227(1)$^b$ & \\
\hline
\hline
\end{tabular}
\begin{flushleft}
Result $^a$ was published in Ref.~\cite{PaSa2000},
for result $^b$ see Ref.~\cite{CeSzJe2001}.
\end{flushleft}
\end{center}
\label{default}
\end{table*}

In view of the above considerations, we can approximate
the black-body frequency in Eq.~\eqref{df_g} to
good approximation by $\omega_B=0$ and evaluate
$g(0)$. This is instructive, because $g(0)$ can be broken down into
distinct contributions, which allows us to present them
separately, possibly enabling an independent verification
of the calculations if needed. We thus calculate first
the quantity
$\alpha_P = \langle \psi_0 | r^i  {\cal R}'(0) r^i | \psi_0 \rangle$,
which is directly connected to the dipole
polarizability by the relation $\alpha_d = -2 \alpha_P/3$
in the nonrelativistic approximation.
The resolvent $ {\cal R}'(0)$ can be replaced
effectively by the sum over $P$ states
as in
\begin{equation}
{\cal R}_P(0) = \sum_n
\frac{| \phi_{nP} \rangle \langle \phi_{nP} |}{ E_0 -  E_n } \,,
\end{equation}
where the sum, for our calculation, is only over the
singlet states,
and $n$ is the principal quantum number.
For the ground
state, all contributions from intermediate states
fulfill $E_0 - E_n < 0$. In that case,
the exact representation of the $P$ state component of the
resolvent gives the lowest possible value for the polarizability,
thus leading to a variational principle for the
determination of the second-order polarizability.

For the calculation of the $g$ function,
we also need the first order correction to the wave function,
\begin{align}
\label{deltaPsiPi}
| \delta \psi_P^i \rangle =& \; {\cal R}_P(0) \; r^i | \psi_0 \rangle
\nonumber\\
=& \; \sum_{m=1}^{N_P} v^P_{m}  \bigr[r^i_1  e^{-a_i r_1 - b_i r_2 - c_i
r_{12} } - (r_1 \leftrightarrow r_2) \bigl],
\end{align}
so that the dipole polarizability
\begin{equation}
\label{alphaP}
\alpha_P = \sum_{i=1}^3 \langle \psi_0 | r^i | \delta \psi_P^i \rangle
\end{equation}
can be written in terms of the dipole matrix element
of the reference state wave function and of the perturbation $| \delta \psi_P^i \rangle$.
Variational parameters for $\delta \psi_P$ are generated just as for
the ground state, but the size $N_P = \tfrac32 N_S$ of the basis
of $P$ states is chosen to be larger than $N_S$
used for the generation of the ground state.
With these results at hand, it is then easy
to calculate the other second-order element
\begin{equation}
\label{alphaPP}
\alpha_{PP} =
\sum_{i=1}^3 \langle \psi_0 |  r^i \,  {\cal R}_P^2(0) \, r^i | \psi_0 \rangle =
\langle \delta \psi_P | \delta \psi_P \rangle
\end{equation}
needed for $g(0)$.

The two fourth-order terms that enter $g(0)$ can be expressed as
\begin{equation}
\label{alphaDS}
\alpha_D =  \langle \delta \psi_P^i | r^j | \delta \psi_D^{ij} \rangle \,,
\qquad
\alpha_S = \langle \delta \psi_P^i | r^i | \delta \psi_S \rangle \,,
\end{equation}
where the intermediate $S$ and $D$ states are represented in the form
\begin{subequations}
\begin{align}
& | \delta \psi_S \rangle =
{\cal R}'_S(0) r^i | \delta \psi_P^i \rangle
\nonumber\\
& \; =
\sum_{m=1}^{\bar N_S} v^S_m \bigr[ e^{-a_i r_1 - b_i r_2 - c_i r_{12} } -
(r_1 \leftrightarrow r_2) \bigl] \,,
\\[2ex]
& | \delta \psi_D^{jk} \rangle =
{\cal R}_D(0) r^j | \delta \psi_P^k \rangle
= \sum_{m=1}^{4 N_D/5} v^D_m
\bigl[ (r_1^j r_1^k - \delta^{jk} r_1^2)
\nonumber\\
& \times
\ee^{-a_i r_1 - b_i r_2 - c_i r_{12} }
- (r_1 \! \leftrightarrow \! r_2)  \bigr]
+ \sum_{m=1}^{N_D/5} \!\! v^{PP}_m
\\
& \times
\bigl[( r_{12}^j r_{12}^k - \delta^{jk} r_{12}^2)
\ee^{-a_i r_1 - b_i r_2 - c_i r_{12} }
- (r_1 \! \leftrightarrow \! r_2) \bigr] \,.
\nonumber
\end{align}
\end{subequations}
The dominant effect due to intermediate $D$ states comes from
the excitation of one of the electrons, and the second case adds a mixture of single
excitations of two electrons.
For the calculation, we choose $2 \bar N_S = 2 N_D = 4 N_S = \tfrac83 \, N_P$.

With the results for $\alpha_P$,
$\alpha_{PP}$, $\alpha_D$ and $\alpha_S$
as defined in Eqs.~\eqref{alphaP}, \eqref{alphaPP} and~\eqref{alphaDS},
(see also Table~\ref{table1}), we can proceed to the
evaluation of
\begin{equation}
\label{df_g0}
\gamma_2 \equiv
g(0) = -\frac23 \, (5 \, \alpha_S + 6\, \alpha_D - 3\, \alpha_P \,  \alpha_{PP} ) \,.
\end{equation}
The prefactors are determined by angular algebra~\cite{VaMoKh1988}.
Here, the term $ \alpha_P \,  \alpha_{PP}$ is equal to the sum
of the last two products of
terms on the right-hand side of Eq.~\eqref{df_g},
and $ \alpha_D$ and $ \alpha_S$ correspond to
the $D$  and $S$ state components of the sum of the first
four terms on the right-hand side of Eq.~\eqref{df_g},
in the limit $\omega_B \to 0$.
Indeed, the two-color hyperpolarizability $\gamma_2 = g(0)$ can be written in terms of
just the matrix elements $ \alpha_D$, $ \alpha_S$,
$ \alpha_P$ and $ \alpha_{PP}$ because we set $\omega_B \to 0$
in Eq.~\eqref{df_g}.

The two-color hyperpolarizability $\gamma_2 = g(0)$,
calculated here, is not to be confused
with the static hyperpolarizability $ \gamma$
which is used in order to describe the
fourth order perturbation when all four photons are from the same
field~\cite{CeSzJe2001,BiPi1989}. The latter can be expressed in our parameters
as
\begin{equation}
\label{defgamma}
\gamma =
-\frac{8}{15} (5 \, \alpha_S + 6 \, \alpha_D - 5 \,  \alpha_P \,\alpha_{PP}) \,.
\end{equation}
The numerical prefactors are different from the ones in Eq.~\eqref{df_g0}.
Using our numerical algorithm, we may verify the
result for the static hyperpolarizability $ \gamma$
given in Ref.~\cite{CeSzJe2001} and add one more
significant digit, $\gamma = 43.104\,227\,7(1)$ as opposed to
$\gamma = 43.104\,227(1)$ from Ref.~\cite{CeSzJe2001} (see also Table \ref{table1}).

\begin{figure}[th!]
\begin{center}
\includegraphics[width=0.8\linewidth]{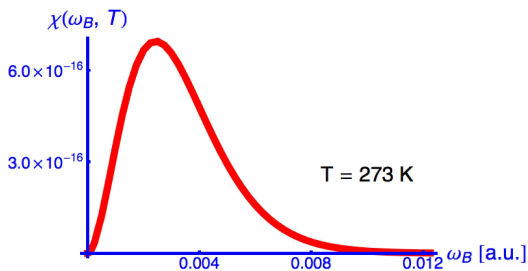}
\caption{\label{fig1} (Color online) $\chi(\omega_B,T)$ as a function of
$\omega_B$ at $T=273$ K. The frequency scale is in
atomic units (a.u.).}
\end{center}
\end{figure}

\begin{figure}[th!]
\begin{center}
\includegraphics[width=0.8\linewidth]{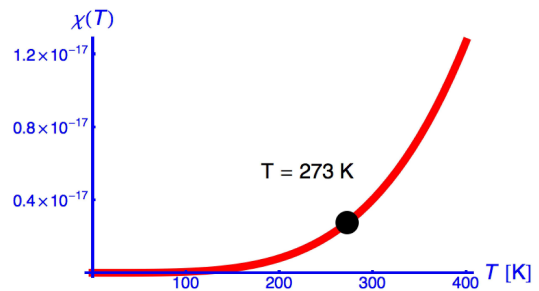}
\caption{\label{fig2}  (Color online)
A plot of the dimensionless quantity
$\chi$ against the temperature $T$
measured in Kelvin (K) is shown, with special
emphasis on the point $T = 273 \, {\rm K}$.}
\end{center}
\end{figure}

The above calculations illustrate the computational procedure for
the static one-color hyperpolarizabilty $\gamma$
and its two-color generalization $\gamma_2$.
For finite excitation frequency, there is a subtlety which deserves
some extra considerations. In the fourth-order matrix elements
in Eq.~\eqref{df_g0}, the outer $S$ state couples to odd parity $P$
states by dipole transitions. The inner virtual states can be
$S$ or $D$ states, but they can also be even parity $P$ states.
Let us suppose that the reference $S$ state has been coupled
to a $z$ polarized odd-parity $P$ state. By a dipole transition,
emitting or absorbing an $y$ polarized photon, the two-electron
system may couple to an even parity $P$ virtual state that
is proportional to the $x$ component of the cross product
$\vec r_1 \times \vec r_2$. For $\omega_B \to 0$, the contribution
of even parity $P$ states vanishes due to symmetry, but it
gives a finite effect for nonvanishing $\omega_B$.
In order to fix the notation, we define
\begin{equation}
\alpha_{PE} =
\epsilon_{ijk} \langle \delta \psi_P^i | r^j | \delta \psi_{PE}^{k} \rangle \,,
\end{equation}
where the $P$ even ($PE$) states are given by
(we invoke the summation convention over $k$ and $l$)
\begin{align}
& | \delta \psi_{PE}^{j} \rangle = \epsilon_{jkl} \,
{\cal R}_{PE}(0) r^k | \delta \psi_P^l \rangle
\\[2ex]
& \; = \sum_{m=1}^{N_{PE}}
v^{PE}_m \bigl[ \epsilon_{jkl}
r_1^k r_2^l \ee^{-a_i r_1 - b_i r_2 - c_i r_{12}} -
(r_1 \leftrightarrow r_2) \bigl] \,.
\nonumber
\end{align}
We choose $N_{PE} = N_S$ and recall the definition of the 
perturbation $| \delta \psi_P^i \rangle$ due to the 
$i$th Cartesian component of the position operator.
The definition of $\alpha_{PE}$ thus involves a total 
of three resolvents of the helium atom.
By numerical calculation, using the same
incremental values for the basis sets as those indicated in Table~\ref{table1},
we obtain $\alpha_{PE} = -0.062\,951\,884\,22(4)$.
This value should not be understood as the static 
value of a dynamic forth-order polarizability
due to even-parity $P$ states (indeed, as outlined above,
the contribution of even-parity $P$ states 
to the two-color hyperpolarizability vanishes).
In our definition of $\alpha_{PE}$,
we have arranged for the $\epsilon$ tensors to isolate 
a nonvanishing contribution in the static limit.
This approach serves two purposes:
(i) to implicitly define the variational parameters used in the 
calculation of the resolvent ${\cal R}_{PE}$ in both the 
static as well as the dynamic regime,
and (ii) to provide a reference value for 
the contribution of even-parity $P$ states to 
a fourth-order matrix element that has a manifestly 
nonvanishing static limit (but no direct physical 
interpretation).

For the integration in Eq.~\eqref{df_chia} over $\omega_B$,
a 120 point Gauss-Legendre quadrature in the interval
$(0,\omega_{B\mathrm{cut}})$ is fully sufficient to obtain
\begin{equation}
\label{chiexact}
\chi = 2.742\,57(1) \times 10^{-18} \qquad \mbox{for} \qquad
T = 273 \, {\rm K} \,.
\end{equation}
Under the approximation $g(\omega_B) \approx g(0)$
in Eq.~\eqref{df_chia},
which we can do because $g$ varies slowly on the frequency scale
of the Boltzmann distribution at room temperature,
the integral over $\omega_B$ can by done analytically. It has a
character similar to the Stefan-Boltzmann law with a $T^4$
temperature dependence. For clarity, we now
return to SI MKSA units.
With the atomic unit quantities
$\overline \alpha_d = 1.383192\dots$ and
$\overline g(0) = \overline\gamma_2 = 59.752019\dots$
defined as in Eq.~\eqref{atu}, we obtain
\begin{align}
\label{analytic}
\chi \approx & \;
\frac{4 \alpha^3}{135}  \, \pi^2 \,
\frac{\overline \gamma_2}{\overline \alpha_d}
\left( \frac{k_B T}{E_h} \right)^4
\nonumber\\
\approx & \; 4.9372 \times 10^{-28} \; \left( \frac{T}{\mathrm{K}} \right)^4 \,.
\end{align}
This approximation gives $\chi = 2.7424 \times 10^{-18} $
for $T = 273\,\KK$
with an error of order $10^{-4}$ compared to Eq.~\eqref{chiexact}.
At room temperature ($T = 300\,\KK$),
we obtain $\chi = 3.999 \times 10^{-18}$.
Figures~\ref{fig1} and~\ref{fig2} illustrate the
Boltzmann weight of the integrand defining $\chi$
and the overall dependence of $\chi$ on the temperature.
As indicated in Eq.~(\ref{defchi}), $\chi$ is the
multiplicative factor that gives the temperature-dependent effective
static polarizability of helium in the presence of the black-body
radiation.

%
%

\section{Conclusions}
\label{conclu}

We have investigated the effect of black-body radiation
on the determination of the helium molar polarizability by
fourth-order perturbation theory in the
quantized electromagnetic field of the probing microwave and
the black-body radiation field.
This shift of the molar polarizability is of interest
because it may affect a conceivable definition of
thermodynamic temperature or a determination of the Boltzmann
constant based on the measurement~\cite{ScGaMaMo2007}.
Indeed, it has been suggested in Ref.~\cite{HoTr1994} that the relative correction to
the polarizability amounts to a
correction as large as $2.1 \times 10^{-10} \, {\rm K}^{-2} \, T^2 = 16 \times
10^{-6}$ at $T = 273 \, \KK$, which would have been
significant compared to the $9.1 \times 10^{-6}$ relative
uncertainty of the measurement \cite{ScGaMaMo2007}.
However, our fourth-order calculations indicate that the effect of black-body
radiation on the measurement of the polarizability of helium reported in
\cite{ScGaMaMo2007} is negligible. In fact, the result of that measurement,
$(A_{\epsilon,\mathrm{meas}}  - A_{\epsilon,\mathrm{theory}})/
A_{\epsilon,\mathrm{theory}} = (1.8 \pm 9.1) \times 10^{-6}$, supports this
conclusion.
The calculation in \cite{HoTr1994}
considers only the frequency dependence, discussed in Ref.~\cite{BhDr1998} for
example, of the interaction of the black-body radiation with the atoms. In
lowest order, this dependence does not affect the interaction of the microwave
radiation with the atoms.

Here, we have presented a fully quantized approach to the calculation of the
black-body radiation correction to the polarizability of helium, and we have
evaluated all expressions numerically [see Eq.~\eqref{chiexact}] as well as
within a semi-analytic approach [see Eq.~\eqref{analytic}].  However, even
without this formalism, an estimate for the effect of the black-body radiation
on the polarizability of helium at room temperature could have been performed
based on known literature references, as follows.  We start from the
ground-state blackbody shift $\delta\nu_0 = 0.11 \,\Hz$ given by Farley and
Wing~\cite{FaWi1981}.  (This value takes into account the 300\,K to
273\,K temperature difference as compared to Ref.~\cite{FaWi1981}.)
This amounts to a relative
change in the polarizability of order $\delta\nu_0/\nu_{\mathrm{He}} \approx
2\times 10^{-17}$, which affects the virtual states in the defining expression
for the ground state polarizability (we denote by $\nu_{\mathrm{He}} \approx 5
\times 10^{15}\, \Hz$ a typical transition frequency in helium).  Unlike the
QED corrections, the black-body radiation effect is largest for highly-excited
states, so a possibly larger correction would come from the shift of the
excited-state energies due to the black-body radiation.  These states enter the
expression for the polarizability as virtual states, and the relative frequency
of the virtual transitions leads to a proportional shift of the polarizability,
as implied by the fourth-order effect given in Eq.~\eqref{df_g}.  Farley and
Wing find that the correction $\delta\nu_i$ for each excited state $i$
approaches a limiting shift of $\delta\nu_\infty \to 2.0 \,\kHz$ as $i \to
\infty$~\cite{FaWi1981}.  For lower states, this gives an overestimate because
the black-body shifts for typical excited states are larger.  Nevertheless, if
all excited state energies were shifted by this amount, then the polarizability
would undergo a relative change of order $\chi(T = 273 \, {\rm K}) =
\delta\nu_\infty/\nu_{\mathrm{He}} \approx 4 \times 10^{-13}$, which needs to
be contrasted to our exact result (the latter is in the range of $10^{-18}$
and confirms the overestimation).

In this article, we have carried out a detailed analysis of the effect, and 
obtained the
precise shift of the polarizability in fourth-order perturbation theory, which
can be expressed as a relative perturbation of the second order polarizability
effect proportional to a dimensionless factor $\chi$.  Our analytic formula
given in Eq.~\eqref{analytic} and our numerical results obtained by numerically
integrating the black-body spectrum confirm this estimate under the proviso that
the shift of the lowest excited helium states, which is of the order of a few
hertz, gives the relevant contribution to the polarizability.  In particular,
we find a shift of $\chi(T= 273\,{\rm K})= 2.7424 \times 10^{-18}$ for the
relative correction to the polarizability of the helium ground state.
Furthermore, according to Eq.~\eqref{analytic}, we find a $T^4$ dependence for
the overall shift of the polarizability with an analytic dependence of the form
$\chi \approx 4.9372 \times 10^{-28} (T/\KK)^4$.  Although we have carried out
numerical calculations only up to a temperature $T = 400\, {\rm K}$, an
analytic estimate based on the known form of the black-body spectrum shows that
the formula~Eq.~\eqref{analytic} should be valid to better than $1 \, \%$ in a
temperature range to at least 2000\,K.

Finally, we would also like to briefly discuss the size of the effect for other
noble gases like neon and argon, which are of experimental interest. The
two-color hyperpolarizability scales as $(Z_{\rm eff} \, \alpha)^{-10}$ where
$Z_{\rm eff}$ is the effective nuclear charge seen by the electron in the
valence shell. The final result for the shift of the polarizability due to the
black-body radiation interaction is obtained as an integral over the dynamic
two-color hyperpolarizability, and its value may thus additionally depend on
the overlap of the resonance frequencies with the maximum of the
temperature-dependent black-body spectrum. For temperatures not exceeding
$400\,{\rm K}$, however,  the {\em static} two-color hyperpolarizability of the
noble gas in question should provide a good estimate of the total effect,
according to Eq.~\eqref{analytic}. We thus expect that the result for $\chi$ in
the temperature range $0 < T < 400\, {\rm K}$ for neon and argon should not
differ from the result given in Eq.~\eqref{analytic} by more than two orders of
magnitude and thus be negligible on the level of accuracy reached in the
experiment~\cite{ScGaMaMo2007} . More accurate estimates require an explicit
calculation of $\gamma_2$ for the atomic reference system under investigation.

%
%

\section*{Acknowledgments}

The authors thank J. W. Schmidt
and M. R. Moldover for helpful discussions.
This work has been supported by the National Science Foundation
and by a precision measurement grant from the
National Institute of Standards and Technology.

\appendix

%
%
\section{Two-electron integrals}
\label{appa}

The two-electron integral $\Gamma$ is defined by
\begin{align}
& \Gamma(n_1,n_2,n_3,\alpha,\beta,\gamma)
\nonumber\\
& \equiv
\int\frac{\dd^3r_1}{4\,\pi}\,
\int\frac{\dd^3r_2}{4\,\pi}\,
\ee^{-\alpha\,r_1-\beta\,r_2-\gamma\,r_{12}}\,
r_1^{n_1-1}\,r_2^{n_2-1}\,r_{12}^{n_3-1}.
\end{align}
This integral takes a very simple form when all $n_i=0$,
\begin{equation}
\Gamma(0,0,0,\alpha,\beta,\gamma) =
\frac{1}{(\alpha+\beta)\,(\alpha+\gamma)\,(\beta+\gamma)}.
\end{equation}
The explicit form for $n_i>0$ can be obtained by differentiation
with respect to the corresponding parameter. For the actual evaluation of $\Gamma$,
we use compact recurrence relations
from Refs.~\cite{SaRoKo1967,Ko2000}.

\end{document}